\documentclass[twocolumn,showpacs,pre,floatfix,superscriptaddress]{revtex4}
\usepackage{graphicx,bm,url}

\newcommand{\un}[1]{\ensuremath{\,\mathrm{#1}}}
\begin{document}

\title{Increment definitions for scale dependent analysis of
  stochastic data} 

\author{Matthias Waechter} \email{matthias.waechter@uni-oldenburg.de}
\affiliation{Institute of physics, Carl-von-Ossietzky University, 
  D-26111 Oldenburg, Germany}  
\author{Alexei Kouzmitchev} 
\affiliation{Institute of theoretical physics, University of M{\"u}nster,
  D-48149 M{\"u}nster, Germany}
\email{kuz@uni-muenster.de} 
\author{Joachim Peinke}
\email{peinke@uni-oldenburg.de} 
\affiliation{Institute of physics, Carl-von-Ossietzky University, 
  D-26111 Oldenburg, Germany}

\date{\today}

\begin{abstract}
  It is common for scale-dependent analysis of stochastic data to use the
  increment $\Delta(t,r) = \xi(t+r) - \xi(t)$ of a data set $\xi(t)$ as a
  stochastic measure, where $r$ denotes the scale.  For joint statistics of
  $\Delta(t,r)$ and $\Delta(t,r')$ the question how to nest the increments on
  different scales $r,r'$ is investigated.  Here we show that in some cases
  spurious correlations between scales can be introduced by the common
  left-justified definition. The consequences for a Markov process are
  discussed. These spurious correlations can be avoided by an appropriate
  nesting of increments.  We demonstrate this effect for different data sets
  and show how it can be detected and quantified. The problem allows to
  propose a unique method to distinguish between experimental data generated by
  a noiselike or a Langevin-like random-walk process, respectively.
\end{abstract}


\maketitle

\section{Introduction}

The complexity of most disordered systems depends on the scale at which they
are observed.
Therefore, stochastic analysis of those systems uses scale-dependent quantities
for their characterization.
The term ``scale'' here means for a  data set $\xi(t)$ the distance $r$ 
between two arbitrary points $t,t'$ with $t'-t=r$ ($t$ may denote
time as well as space in this context).
The \emph{increment} $\Delta(t,r) = \xi(t+r) - \xi(t)$ is a common
scale-dependent measure of complexity and disorder.
Well-known examples for other scale-dependent measures of complexity are the
autocorrelation function $R(r)=\langle\xi(t)\xi(t+r)\rangle$, the rms width
$w_r(t)=\langle[\xi(t)-\bar\xi]^2\rangle^{1/2}_r$, or wavelet functions.

Traditionally, the investigation of statistical properties is performed on
distinct scales, e.g., by means of the structure functions
$\langle\Delta(t,r)^n\rangle$ given by the probability density functions (PDF)
$p(\Delta(t,r))$. An advanced approach is to try to describe the joint
statistics of the chosen measure on many different scales.  This is achieved by
the knowledge of the joint PDF $p(\Delta(t,r_1); \ldots;\Delta(t,r_n))$. By
these joint PDF also the correlations between scales are worked out, showing
how the complexity is linked between scales.

If the statistics of the scale-dependent measure can be regarded as a Markov
process evolving in $r$, the knowledge of two-scale conditional PDF
is sufficient for a complete description of multiscale joint PDF
\cite{Risken1984}.
The conditional PDF $p(\Delta_1(t,r_1)|\Delta_0(t,r_0))$ denotes the
probability of finding an increment $\Delta(t,r_1)=\Delta_1$ on the scale
$r_1$ under the condition that at the same time $t$ on a different scale $r_0$
another increment $\Delta(t,r_0)=\Delta_0$ has been found.
The validity of the Markov property can be tested by the
investigation of conditional PDF \cite{Renner2001}, of the Chapman-Kolmogorov
equation \cite{Friedrich1997a}, or of reconstructed noise \cite{Marcq2001}.
If, furthermore, the
noise involved in the process is Gaussian distributed, the whole joint
statistics can be grasped by a Fokker-Planck or Langevin equation
\cite{Risken1984,Friedrich1997a,Renner2001,Waechter2003,Marcq2001}.
This approach has been used by different researchers in a number of
applications
\cite{Friedrich1997a,Renner2001,Waechter2003,Naert1997,Marcq2001,Jafari2003,
  Ghasemi2003,Ausloos2003}. In some cases also the question of increment
definitions has been discussed, as in \cite{Naert1997}.

In this paper we want to address the question of whether the relative location
of the increments may introduce spurious correlations between different
scales. In particular, we investigate the two cases of the left-justified
increment
\begin{equation}
  \label{eq:increments_incr_left}
  \Delta_l(t,r) = \xi(t+r) - \xi(t)
\end{equation}
and the centered increment
\begin{equation}
  \label{eq:increments_incr_centered}
 \Delta_c(t,r) = \xi(t+r/2) - \xi(t-r/2) \;.
\end{equation}
We will discuss the implications of these increment definitions on two
different types of stochastic processes and compare the results for
experimental data.

\newlength{\breite}
\setlength{\breite}{0.33\textwidth}
\begin{figure*}[htbp]
  \begin{center}
    \hspace*{\fill}
    \includegraphics[width=\breite]{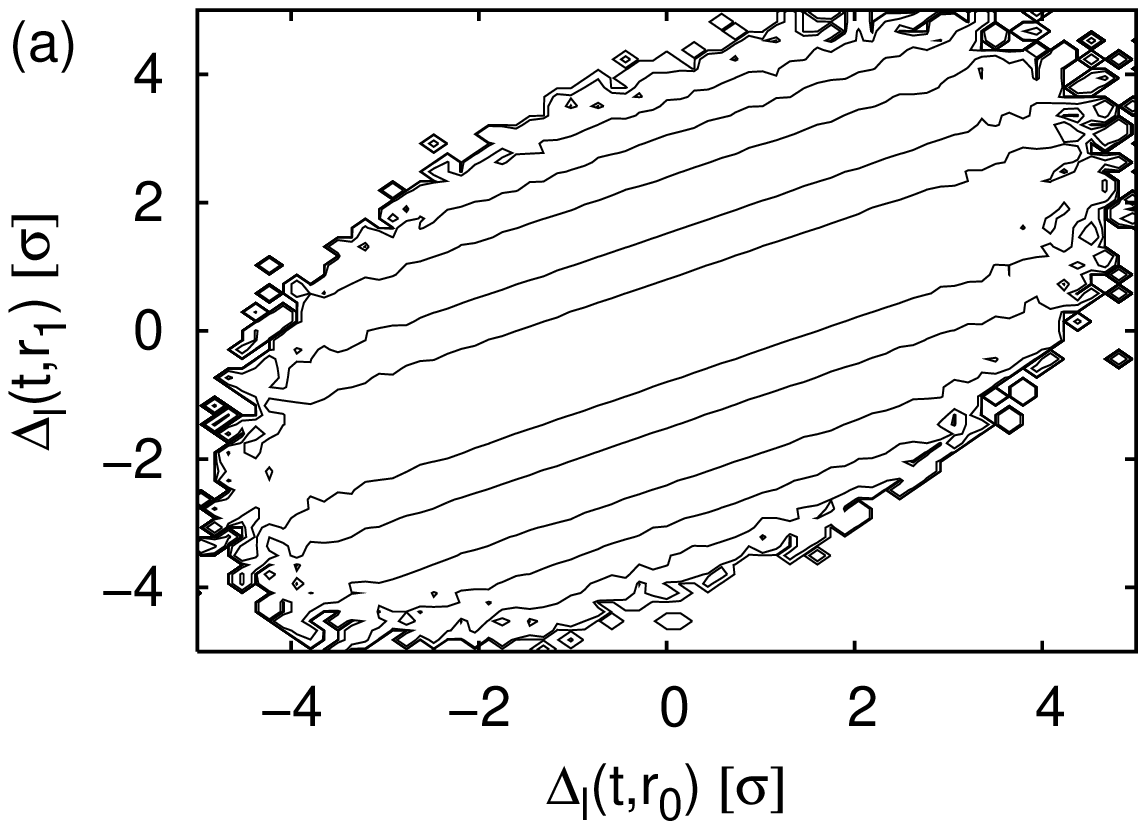}
    \hspace*{\fill}
    \includegraphics[width=\breite]{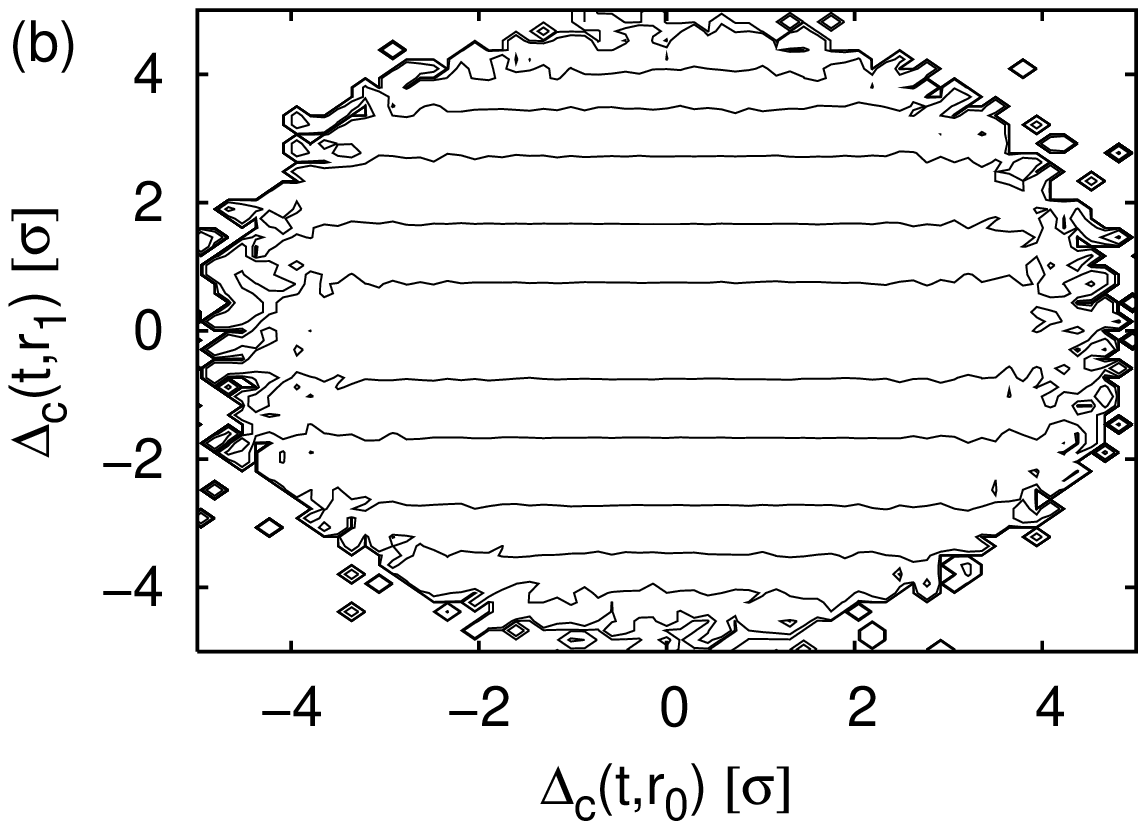}
    \hspace*{\fill}
    \caption[corr.\ and uncorr.\ condpdfs]{
      Conditional PDF of (a) left-justified and (b) centered increments of
      a white-noise-like random walk.
      Scales $r_0$ and $r_1$ are 116 and 100
      sample steps, respectively. PDF are displayed as contour lines, the
      levels differ by a factor of 10, with an additional level at 0.3.  }
    \label{fig:increments_gauss}
  \end{center}
\end{figure*}

\section{Increment definitions and correlations between scales}

First, two idealized types of stochastic processes for
discrete times are defined.
The first one is called white-noise-like random walk (WNR) with the random
variable $\xi(t_i)$ given by 
\renewcommand{\arraystretch}{1.5}
\begin{equation}\label{eq:increments_mod1} 
  \begin{array}{rcl}
    P(\xi,t_i)&=&\frac{1}{\sqrt{2 \pi \sigma^2}}
    \exp\left\{-\xi^2/(2\sigma^2)\right\}\;, \\
    P(\xi,t_i;\xi',t_k)&=&P(\xi,t_i) P(\xi',t_k) 
    \quad\mbox{for $i \neq k$},
  \end{array}
\end{equation} 
which implies that 
\begin{eqnarray}
  \label{eq:increments_mod1_imp1}
  \left<\xi(t_i) \xi(t_k)\right> &=& \sigma^2 \delta_{ik} \\
  \mbox{and}\qquad
  \left<\xi(t_k)| \xi(t_i)=x_0\right> &=& 0   
  \quad\mbox{for $k>i$.}
  \label{eq:increments_mod1_imp2}
\end{eqnarray}
The second model, called Langevin-like random walk (LRW), 
is the cumulative sum of the first one, i.e.,
\begin{equation} \label{eq:increments_mod2} 
  \begin{array}{c}
    \xi(t_i)=\sum_{k=1}^i w(t_k) \quad,
  \end{array}
\end{equation} 
with $w(t_k)$ distributed as in (\ref{eq:increments_mod1}).
This implies that here,
\begin{eqnarray}
  \label{eq:increments_mod2_imp1}
  \left<\xi(t_i) \xi(t_k)\right> &=& i \, \sigma^2 
  \quad\mbox{for $k \ge i$}\\
  \mbox{and}\quad
  \left<\xi(t_k)| \xi(t_i)=x_0\right> &=& x_0  
  \quad\mbox{for $k>i$.}
  \label{eq:increments_mod2_imp2}
\end{eqnarray}

The correlations between different scales for both increment functions
$\Delta_l(t,r)$ and $\Delta_c(t,r)$ can be calculated easily. Taking increments
of the WNR (\ref{eq:increments_mod1}) and $r_1>r_0>0$ we get
\begin{equation}
  \label{eq:increments_mod1_icorr}
  \left< \Delta_l(t,r_0)\Delta_l(t,r_1)\right> = \sigma^2 \;, \;
  \left< \Delta_c(t,r_0)\Delta_c(t,r_1)\right> = 0 \;.
\end{equation}
Note that for arbitrary different scales of $\Delta_l$ correlations are
present even though there are no correlations at all in the data $\xi(t)$.
These spurious correlations are introduced by the left-justified increment
because for a fixed value $t$ the increments $\Delta_l(t,r)$ of all scales $r$
have the term $\xi(t)$ in common, see Eq.~(\ref{eq:increments_incr_left}).
For the LRW we obtain
\begin{equation}
  \label{eq:increments_mod2_icorr}
  \left< \Delta_l(t,r_0) \Delta_l(t,r_1)\right> =  
  \left< \Delta_c(t,r_0) \Delta_c(t,r_1)\right> = r_0\sigma^2
\end{equation}
with $r_1>r_0>0$.
In contrast to the WNR, identical correlations of both increment
definitions result here. 

Correlations between two scales $r_0,r_1$ can
directly be observed as dependence of $p(\Delta_1|\Delta_0)$ on $\Delta_0$.
Increment statistics of the WNR
are presented in Fig.~\ref{fig:increments_gauss} as conditional pdf
of left-justified and centered increments.
Data have been generated using the routine gasdev from \cite{Press1992} and
normalized by $\sigma$.
As expected from Eq.~(\ref{eq:increments_mod1_icorr}), for the \emph{left-justified}
increments (a) a correlation between both scales $r_1,r_0$ is evident because
$p(\Delta_l(t,r_1)|\Delta_l(t,r_0))$ strongly depends on the value of
$\Delta_l(t,r_0)$. In contrast, the conditional PDF of the \emph{centered}
increments (b) is independent of $\Delta_c(t,r_0)$ and thus both scales are
uncorrelated for centered increments.
For corresponding diagrams calculated for the LRW process (not shown
here) we find identical PDF for left-justified and centered increments, similar
to Fig.~\ref{fig:increments_gauss}(a), and in accordance with Eq.~(\ref{eq:increments_mod2_icorr}).

\section{Consequences for Markov properties}
\label{sec:increments_Markov}

\begin{figure*}[htbp]
  \begin{center}
    \includegraphics[width=\breite]{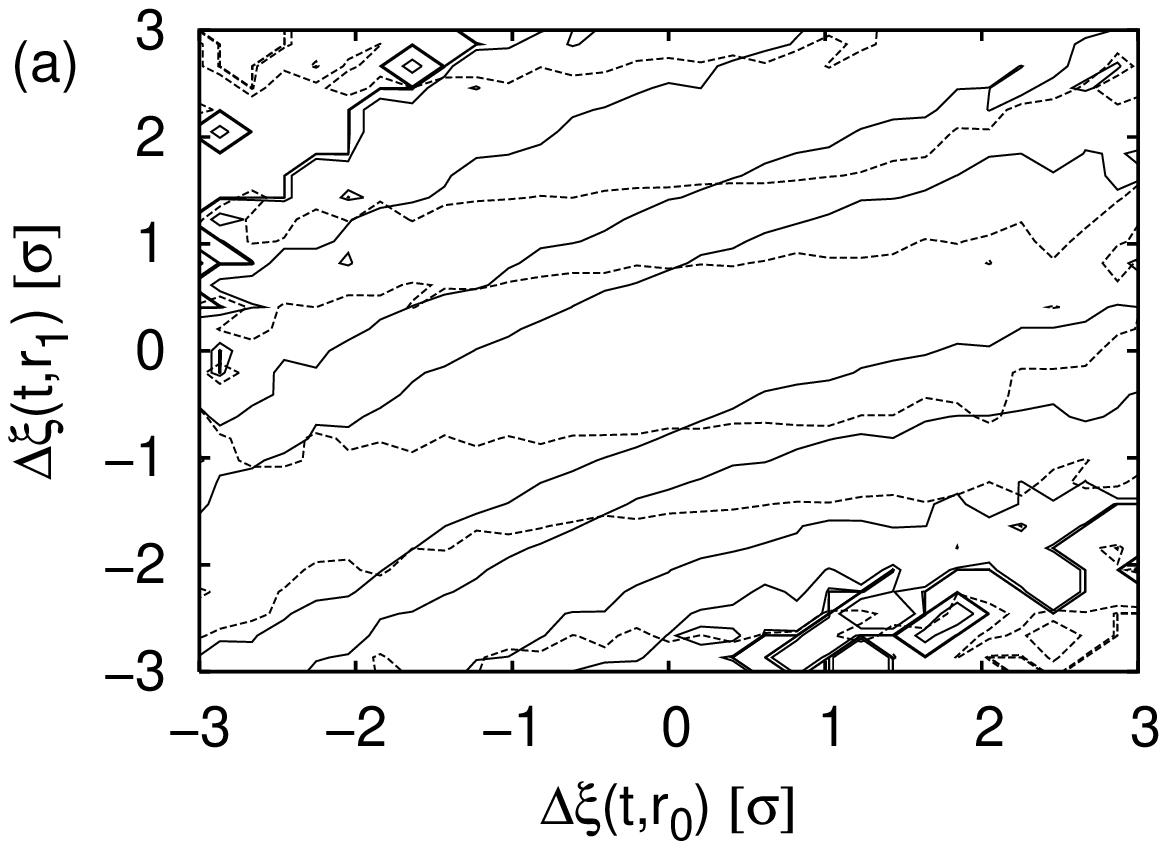}%
    \includegraphics[width=\breite]{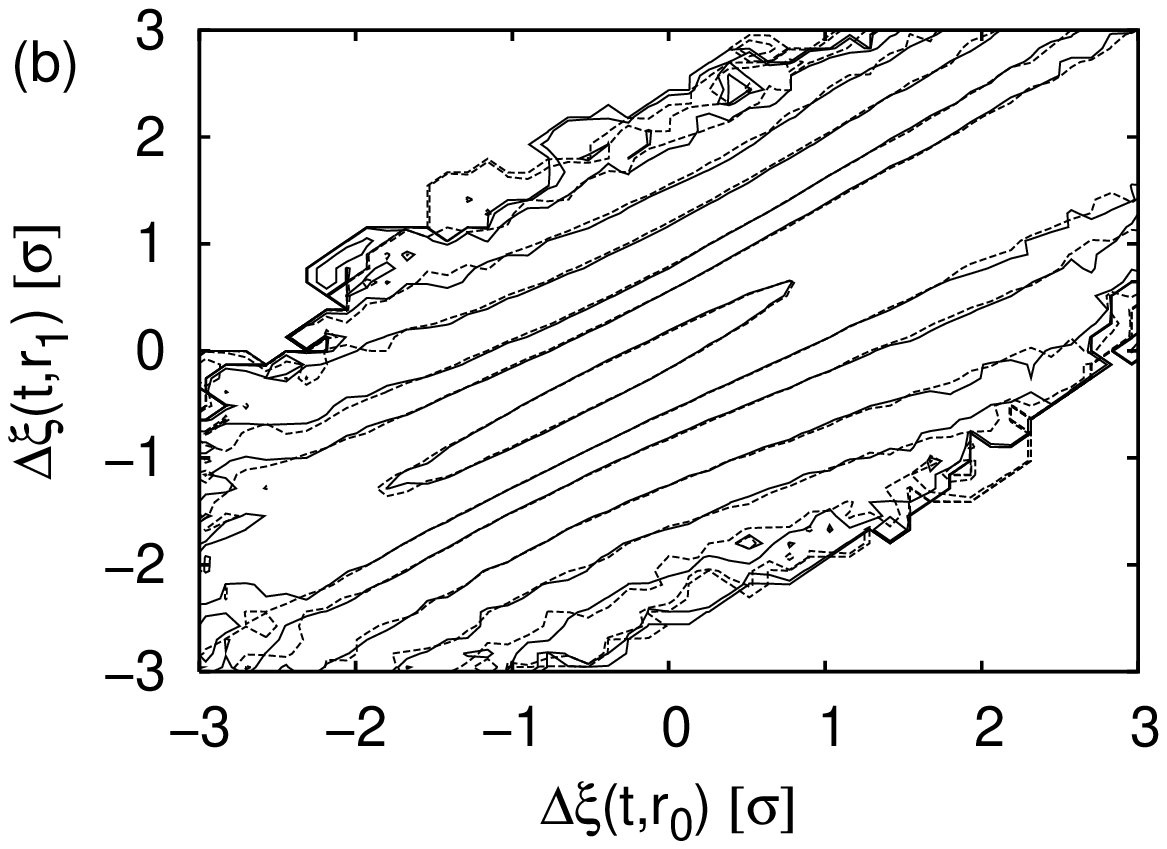}%
    \includegraphics[width=\breite]{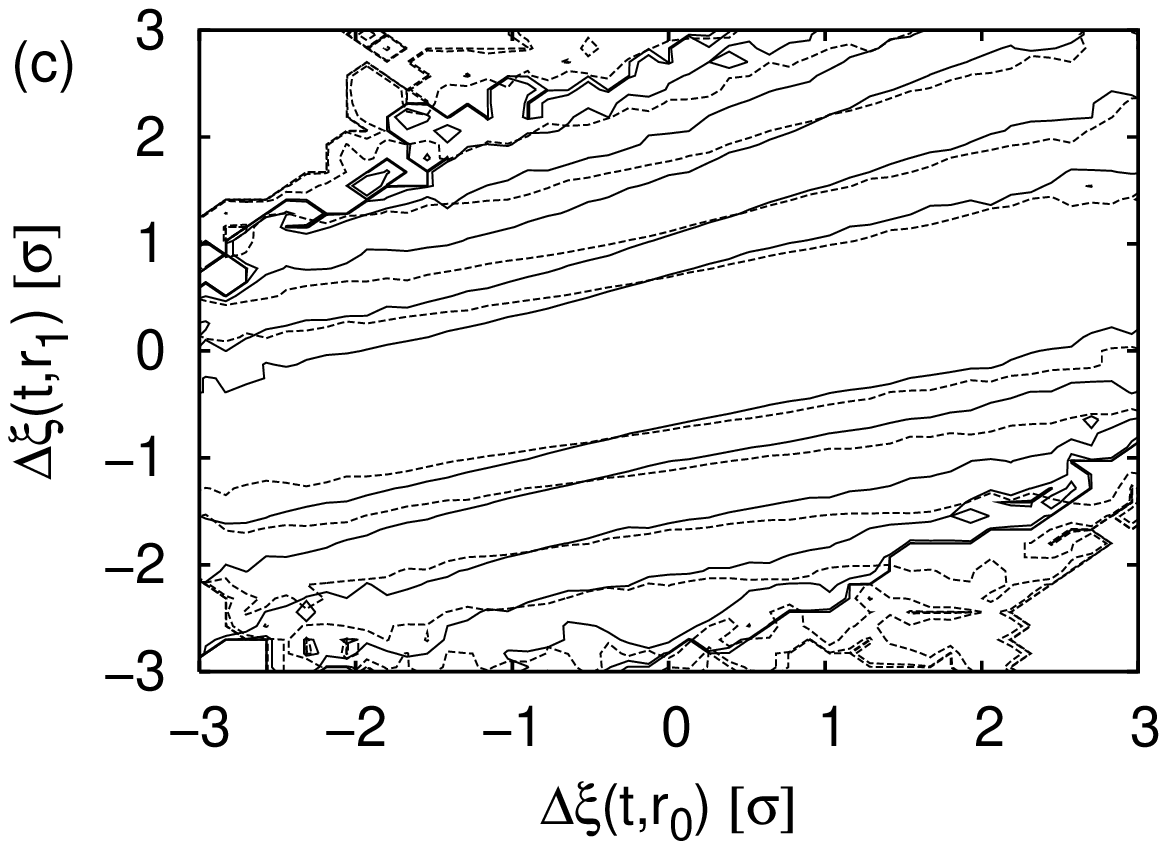}
    \caption[Condpdfs Asphalt \& Turbulenz]{
      Conditional PDF of left-justified (solid lines) and centered (broken
      lines) increments of experimental data. (a) Height profiles from a
      smooth asphalt road using scales $r_0=137,\, r_1=104$\,mm , (b,c)
      velocity time series from a turbulent free jet, with small (b) and large
      (c) scale differences $\delta r=r_0-r_1$ (see text). PDF are plotted as
      in Fig.~\ref{fig:increments_gauss}.  }
    \label{fig:increments_Condpdfs}
  \end{center}
\end{figure*}

For the analysis and description of stochastic data by means of a
Fokker-Planck equation, as mentioned in the Introduction, the underlying
process has to be Markovian.  In the previous section we have shown that the
left-justified increment definition can introduce additional correlations.
This effect also influences the Markov properties of these increments.

It is indeed straightforward to see that the left-justified increment
$\Delta_l(t,r)$ of a WNR is consequently not a Markov process in the scale
variable $r$.
A necessary condition for a stochastic process to be Markovian is
the Chapman-Kolmogorov equation \cite{Risken1984} (here we use the notation 
$\Delta_l(t,r_i)=\Delta_{l,i}$)
\begin{equation}
  \label{eq:increments_CKE}
  p(\Delta_{l,3}|\Delta_{l,1}) = \int_{-\infty}^\infty 
  p(\Delta_{l,3}|\Delta_{l,2}) p(\Delta_{l,2}|\Delta_{l,1}) d\Delta_{l,2}
\end{equation}
for any triplet $r_1<r_2<r_3$.

For the WNR process we first derive from (\ref{eq:increments_incr_left}) and
(\ref{eq:increments_mod1}) the correlation matrix $\bm{S}$ for
$\Delta_{l,1},\Delta_{l,2}$ with elements
$s_{ij}=\langle\Delta_{l,i}\Delta_{l,j}\rangle$, $i,j=1,2$:
\begin{equation}\label{eq:increments_corr_mtrx}
  \bm{S}=\left(
    \begin{array}{cc}
      2\sigma^2 &  \sigma^2\\
       \sigma^2 & 2\sigma^2
    \end{array}\right)
\end{equation}

Because the difference of two Gaussian distributed random variables is also
Gaussian, we can derive two-dimensional PDF of $\Delta_{l,1},
\Delta_{l,2}$ using the general two-dimensional form of the Gaussian
distribution (after~\cite{Risken1984})
\begin{equation}
  \label{general_gauss}
  p(\Delta_{l,i};\Delta_{l,j})=\frac{1}{2\pi\sqrt{\det \bm{S}}}
  \exp\left\{-\frac{1}{2}\sum\limits_{i,j=1}^2
    (\bm{S}^{-1})_{ij}\Delta_{l,i}\Delta_{l,j} \right\}
\end{equation}
Using (\ref{eq:increments_corr_mtrx}) and (\ref{general_gauss}), we can now explicitly
calculate both sides of Eq.~(\ref{eq:increments_CKE}), namely
\begin{equation}
  \label{eq:increments_CKE_left}
  p(\Delta_{l,3}|\Delta_{l,1})=\frac{1}{\sqrt{2\pi}\sqrt{3/2}
      \sigma}\exp{-\frac{(\Delta_{l,3}-\Delta_{l,1}/2)^2}{3\sigma^2}}
\end{equation}
and
\begin{eqnarray}
  \lefteqn{
  \int_{-\infty}^{\infty}
  p(\Delta_{l,3}|\Delta_{l,2}) p(\Delta_{l,2}|\Delta_{l,1}) d\Delta_{l,2}=
  }\nonumber\\
  &&
  \frac{1}{\sqrt{2\pi}\sqrt{15/8}\,\sigma}
  \exp{-\frac{(\Delta_{l,3}-\Delta_{l,1}/4)^2}{(15/4)\sigma^2}}\;.
  \label{CKE_right}
\end{eqnarray}
Obviously the Chapman-Kolmogorov equation (\ref{eq:increments_CKE}) is violated for
left-justified increments of the WNR on any scales $r_1<r_2<r_3$. The same
procedure can be used to see that for centered increments the Markov property
holds.

\section{Indicators for spurious correlations caused by increment definition}

The question if, or if not, the above-mentioned spurious correlations between
different scales are introduced by the increment definition is of
practical importance for the analysis of measured data.
For data which behave like the LRW, i.e.,
$ \langle\xi(t+r)\,|\,\xi(t)\!=\!x_0\rangle = x_0 $ rather than 
$ \langle\xi(t+r)\,|\,\xi(t)\!=\!x_0\rangle = 0 $ as for the WNR, the increment
definition should be unimportant. No spurious correlations would be created in
either case. In contrary, for data which behave more like the WNR, the
increment definition should be more important.

As shown above, the conditional PDF can serve as a means to discriminate
between true and spurious correlations between scales if we compare
conditional PDF of left-justified and centered increments. In
Fig.~\ref{fig:increments_Condpdfs} conditional PDF are shown for these
increments of two experimental data sets.
Figure~\ref{fig:increments_Condpdfs}(a) displays PDF of both increment types
obtained from surface height profiles of a smooth asphalt road. The distance
between consecutive data points is 1.04\un{mm}; further details of the
measurement are found in \cite{Waechter2002a,Waechter2003}. The difference
between both types of increments is evident and similar to that for the WNR in
Fig.~\ref{fig:increments_gauss}.
In Figs.~\ref{fig:increments_Condpdfs}(b) and 2(c) conditional PDF are shown in
the same manner for velocity increments measured in a turbulent free jet at
$\mathit{Re}=2.7\times 10^4$ (for details see \cite{Renner2001}). In both cases
scale $r_1$ is $L/2$. The scale difference $\delta r=r_0-r_1$ is small
($1.5\,\lambda$) for (b) and large ($L$) for (c)
\cite{Integral_and_Taylor_length}.  It can be seen in
Fig.~\ref{fig:increments_Condpdfs}(b) that here conditional PDF of
left-justified and centered increments are identical and the increment
definition does not influence the statistics. For $\delta r=L$ at the end of
the inertial range, Fig.~\ref{fig:increments_Condpdfs}(c), a slight difference
of both conditional PDF has already occurred.
The definition of $L$ \cite{Integral_and_Taylor_length} provides that for
large scale differences $\delta r>L$ a transition to a noiselike behavior as
in figs.~\ref{fig:increments_Condpdfs}(a) and \ref{fig:increments_gauss} can
be expected.
Nevertheless, only a small fraction of the correlation between the scales
$r_0,r_1$ is detected as spurious here.

As a second indicator the conditional expectation value
\begin{equation}
  \label{eq:increments_T1}
  T(r,x_0)=\langle\Delta_l(t,r)\,|\,\xi(t)\!=\!x_0\rangle
\end{equation}
can be estimated from the measured data. $T(r,x_0)$ quantifies the influence
of the value $\xi(t)=x_0$ on the left-justified increment $\Delta_l(t,r)$.  It
follows immediately that $ T(r,x_0) =
\langle\xi(t+r)\,|\,\xi(t)\!=\!x_0\rangle-x_0 $. With eqs.\ 
(\ref{eq:increments_mod1_imp2}) and (\ref{eq:increments_mod2_imp2}) we obtain the ideal cases
$T(r,x_0)=-x_0$ for the WNR and $T(r,x_0)=0$ for
the LRW.
If for experimental data there is a strong dependence of $T$ on $x_0$ the data
must be regarded as noiselike in the sense of the WNR in the
respective length scale, and for scale-dependent analysis the use of
left-justified increments is not appropriate.
If otherwise $T$ is independent of $x_0$ the data behave like 
a LRW, and thus the increment definition is not
important.

\begin{figure}[htbp]
  \begin{center}
    \includegraphics[width=0.45\textwidth]{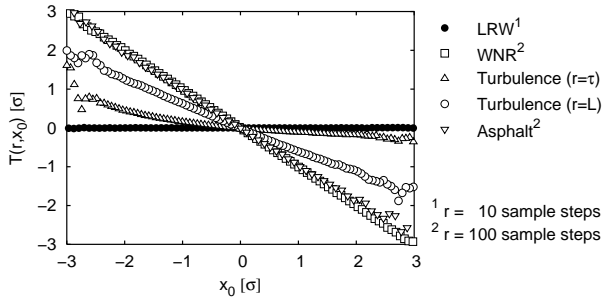}
    \caption{%
      The conditional expectation value
      $T(r,x_0)=\langle\Delta_l(t,r)\,|\,\xi(t)\!=\!x_0\rangle$ for
      different data sets. $T$ is shown as function of $x_0$ for fixed
      scales $r$.
     }
    \label{fig:increments_T1}
  \end{center}
\end{figure}

In Fig.~\ref{fig:increments_T1} we present the dependence of $T$ on
$\xi(t)=x_0$ for different data sets. Data of both ideal cases were generated
as in Fig.~\ref{fig:increments_gauss}. Turbulence and asphalt road data have
already been shown in Fig.~\ref{fig:increments_Condpdfs}.
As expected, we see that for the LRW there is no
dependence of $T(r,x_0)$ on $x_0$. In contrast, for the WNR
as well as for the surface data the dependence is clear with $T(r,x_0)=-x_0$.
For the turbulent velocity increments it can be seen that on the small scale
$\Delta r=\lambda$ the influence of $x_0$ on $T$ is only small, while on the
large scale $\Delta r=L$ the dependence is more pronounced. This finding
corresponds to Fig.~\ref{fig:increments_Condpdfs}, where for small scales
[Fig.~\ref{fig:increments_Condpdfs}(b)] the conditional PDF of left-justified
and centered increments were identical, while for large scales
[Fig.~\ref{fig:increments_Condpdfs}(c)] a difference occurred.

\section{Conclusions}
\label{sec:increments_conclusions}

We found that for scale-dependent analysis of stochastic data, where the
connections between different scales are investigated using increment
statistics, the definition of the increment can be important, depending on the
nature of the data. Apparent correlations between scales may be introduced by
the left-justified increment. The importance of the increment definition
varies between the ideal cases of the LRW (\ref{eq:increments_mod2}), where it
is nonrelevant, and the WNR (\ref{eq:increments_mod1}), where it is
crucial. In this case the use of left-justified increments leads to biased
results for correlations between different scales.
Especially, the surface measurement data we have studied require the centered
definition on all accessible scales \cite{Waechter2003,Waechter2004}. 
For turbulent velocities this influence depends on the regarded length (or
time) scale $r$.
In previous publications \cite{Renner2001,Renner2001Diss} no significant
difference between the drift and diffusion coefficients of the Fokker-Planck
equation of $\Delta_l$ and $\Delta_c$ was found. This is in accordance with
our findings in Fig.~\ref{fig:increments_Condpdfs}(b), where the PDF of
$\Delta_l$ and $\Delta_c$ are shown to be identical for small scale
differences $r_0-r_1$, and only at the integral length scale a difference
occurs [see Fig.~\ref{fig:increments_Condpdfs}(c)].
Detailed consequences are currently being investigated
\cite{Siefert2004}.

The conditional expectation value $T(r,x_0)$ allows to quantify the influence
of a left-justified increment. Nevertheless, the specification of a threshold
in a statistically meaningful way is still an open question.

While in this paper we used the increments (\ref{eq:increments_incr_left}) and
(\ref{eq:increments_incr_centered}) to demonstrate the introduction of spurious
correlations, we expect that these considerations can be applied to general
scale-dependent measures of complexity, such as the rms width
$w_r(t)=\langle(\xi(t)-\bar\xi)^2\rangle^{1/2}_r$ 
or wavelet functions. One could generally distinguish between measures
which are orthogonal on different scales and those which are not
\cite{Orthogonality}.
We expect similar results for correlations between scales as demonstrated here
for left-justified and centered increments.

\acknowledgments
\vspace{-3ex}
%
We experienced helpful discussions with R.~Friedrich, M.~Siefert, M.~Haase, and
A.~Mora. Financial support by the german Volkswagen Foundation is kindly
acknowledged.
%

\bibliography{mw,Increments}

\begin{thebibliography}{16}
\bibitem[{Int()}]{Integral_and_Taylor_length}
\bibinfo{note}{The Taylor length $\lambda$ and the integral length $L$ denote
  lower and upper bound of the so-called inertial range of scales where a
  universal behaviour of turbulent flows is found. $L$ is defined by the
  autocorrelation function and estimates a correlation length such that the
  turbulent fluctuations on scales $r>L$ can be considered as uncorrelated.}
\bibitem[{Ort()}]{Orthogonality}
\bibinfo{note}{Orthogonality can be defined if we construct for $\Delta_l(t,r)$
  a generating function $F_l(t,r,t')=\delta(t+r,t')-\delta(t,t')$ such that
  $\Delta_l(t,r)=\int F_l(t,r,t')\xi(t')\mathrm{d}t'$. Analogously
  $F_c(t,r,t')=\delta(t+r/2,t')-\delta(t-r/2,t')$ is constructed for the
  centered increment $\Delta_c(t,r)$. Now for any $r_0\!\neq\! r_1,\:
  r_0,r_1\!\neq\! 0$ the scalar product $F_l(t,r_0,t')\cdot F_l(t,r_1,t')=\int
  F_l(t,r_0,t')F_l(t,r_1,t')\mathrm{d}t'$ for the left-justified increment is
  obviously different from zero, while $F_c(t,r_0,t')\cdot F_c(t,r_1,t')=0$.}
\end{thebibliography}
\bibliographystyle{apsrev}

\end{document}